\documentclass[pdflatex,sn-mathphys-num]{sn-jnl}% Math and Physical Sciences Numbered Reference Style 
\usepackage{graphicx}%
\usepackage{multirow}%
\usepackage{amsmath,amssymb,amsfonts}%
\usepackage{amsthm}%
\usepackage{mathrsfs}%
\usepackage[title]{appendix}%
\usepackage{xcolor}%
\usepackage{textcomp}%
\usepackage{manyfoot}%
\usepackage{booktabs}%
\usepackage{algorithm}%
\usepackage{algorithmicx}%
\usepackage{algpseudocode}%
\usepackage{listings}%
\usepackage{xr}
\theoremstyle{thmstyleone}%
\theoremstyle{thmstyletwo}%

\theoremstyle{thmstylethree}%

\begin{document}

\title[Article Title]{Optical Neural Engine for Solving Scientific Partial Differential Equations}

%%  \sfx{IV}}\email{iauthor@gmail.com}v
%%=============================================================%%

\author*[1]{Yingheng Tang}\equalcont{These authors contribute equally}\email{ytang4@lbl.gov}

\author[2]{Ruiyang Chen}\equalcont{These authors contribute equally}

\author[2]{Minhan Lou}

\author[2]{Jichao Fan}

\author[3]{\fnm{Cunxi} \sur{Yu}}

\author[1]{\fnm{Andy} \sur{Nonaka}}

\author*[1]{\fnm{Zhi (Jackie)} \sur{Yao}}\email{jackie\_zhiyao@lbl.gov}

\author*[2]{\fnm{Weilu} \sur{Gao}}\email{weilu.gao@utah.edu}

\affil[1]{Lawrence Berkeley National Laboratory, Berkeley, CA 94720, USA}

\affil[2]{Department of Electrical and Computer Engineering, The University of Utah, Salt Lake City, UT 84112, USA}

\affil[3]{Department of Electrical and Computer Engineering, University of Maryland, College Park, MD 20742, USA}


\abstract{Solving partial differential equations (PDEs) is the cornerstone of scientific research and development. Data-driven machine learning (ML) approaches are emerging to accelerate time-consuming and computation-intensive numerical simulations of PDEs. Although optical systems offer high-throughput and energy-efficient ML hardware, there is no demonstration of utilizing them for solving PDEs. Here, we present an optical neural engine (ONE) architecture combining diffractive optical neural networks for Fourier space processing and optical crossbar structures for real space processing to solve time-dependent and time-independent PDEs in diverse disciplines, including Darcy flow equation, the magnetostatic Poisson's equation in demagnetization, the Navier-Stokes equation in incompressible fluid, Maxwell's equations in nanophotonic metasurfaces, and coupled PDEs in a multiphysics system. We numerically and experimentally demonstrate the capability of the ONE architecture, which not only leverages the advantages of high-performance dual-space processing for outperforming traditional PDE solvers and being comparable with state-of-the-art ML models but also can be implemented using optical computing hardware with unique features of low-energy and highly parallel constant-time processing irrespective of model scales and real-time reconfigurability for tackling multiple tasks with the same architecture. The demonstrated architecture offers a versatile and powerful platform for large-scale scientific and engineering computations.}

\maketitle

\section*{Introduction}

Partial differential equations (PDEs) derived from physical laws have been a powerful and faithful computational tool to accelerate the exploration and validation of scientific hypotheses instead of performing expensive and time-consuming real-world experiments~\cite{AzizzadenesheliEtAl2024NRP}. Hence, numerically solving PDEs is essential for scientific research and development in nearly every scientific domain. For example, the interaction of electromagnetic waves with materials and engineered structures in broad applications such as communication, imaging, sensing, and quantum technologies is governed by Maxwell's equations~\cite{GriffithsEtAl2023}; automotive and flight aerodynamics for designing and manufacturing road vehicles and airplanes is determined by Navier-Stokes equation~\cite{BatchelorEtAl2000}; the Earth system including temperature, atmosphere, and ice sheets for understanding climate change and making policies is also described with a series of PDEs~\cite{GoosseEtAl2015}. However, current numerical simulation methods to solve PDEs, such as finite difference/volume methods to solve Maxwell's and the Navier-Stokes equations, are costly in computing time and resources. 

Machine learning (ML) offers a new perspective on solving PDEs through data-driven approaches to enable fast and accurate simulations of many multiphysics and multiscale processes~\cite{JiangEtAl2020NRM,ZobeiryEtAl2021EAAI,VinuesaEtAl2022NCS}. However, the ML model deployment on electronic computing hardware requires substantial computing resources and consumes substantial energy. In the foreseeable future, the fundamental quantum mechanics limit will lead to a bottleneck of further reducing the energy consumption and simultaneously increasing the integration density of electronic circuits to catch up with the increasing scale of ML models in demand for solving complex problems~\cite{TheisEtAl2017CSE,LeisersonEtAl2020S}, thus urgently calling for new high-throughput and energy-efficient ML hardware accelerators. Recently, optical architectures, including photonic integrated circuits for matrix-vector multiplication (MVM)~\cite{ShenEtAl2017NP,FeldmannEtAl2021N}, for neuro-inspired spiking neural networks~\cite{FeldmannEtAl2019N,JhaEtAl2022JLT}, and for photonic reservoir computing~\cite{VandoorneEtAl2014NC,Van-der-SandeEtAl2017N}, and free-space optical systems for MVM~\cite{HamerlyEtAl2019PR,WangEtAl2022NC,ChenEtAl2023NP} and diffractive optical neural networks (DONNs)~\cite{LinEtAl2018S,ZhouEtAl2021NP,ChenEtAl2022LPR,HuEtAl2024NC}, are emerging as high-performance ML hardware accelerators by leveraging different particles -- photons -- to break down electronic bottleneck thanks to high parallelism and low static energy consumption of photons~\cite{WetzsteinEtAl2020N}. However, to date, there is no deployment of any optical computing systems for solving PDEs in any scientific domain.

Here, we present a fully reconfigurable and scalable optical neural engine (ONE) architecture that combines DONN systems for processing data in Fourier space and optical crossbar (XBAR) structures for processing data in real space to solve two-dimensional (2D) spatiotemporal profiles in time-independent and time-dependent PDEs. The ONE architecture not only leverages the advantages of high-performance dual-space processing~\cite{LiEtAl2020APA}, but also can be implemented using optical computing hardware with unique features of low-energy and highly parallel constant-time processing irrespective of model scales, and real-time reconfigurability for tackling multiple tasks with the same architecture. We numerically and experimentally demonstrate the capability of the ONE architecture in solving a broad range of PDEs in diverse disciplines, including the Darcy flow equation in fluid dynamics, the magnetostatic Poisson's equation in micromagnetics, the Navier-Stokes equation in aerodynamics, Maxwell's equations in nanophotonics, and coupled electric current and heat transfer equations in a multiphysics electrical heating problem. The ONE architecture not only outperforms traditional PDE solvers because of its data-driven nature, but also shows comparable and better performance with other ML models while with substantial hardware advantages because of its implementation in the optical domain. The demonstrated ONE architecture is versatile and can be tailored with different combinations of DONN and XBAR structures for solving various PDEs, offering a transformative universal solution for large-scale scientific and engineering computations. 

\section*{Results}
\subsection*{ONE Architecture}

\begin{figure}[hbt]
    \centering
    \includegraphics[width=\textwidth]{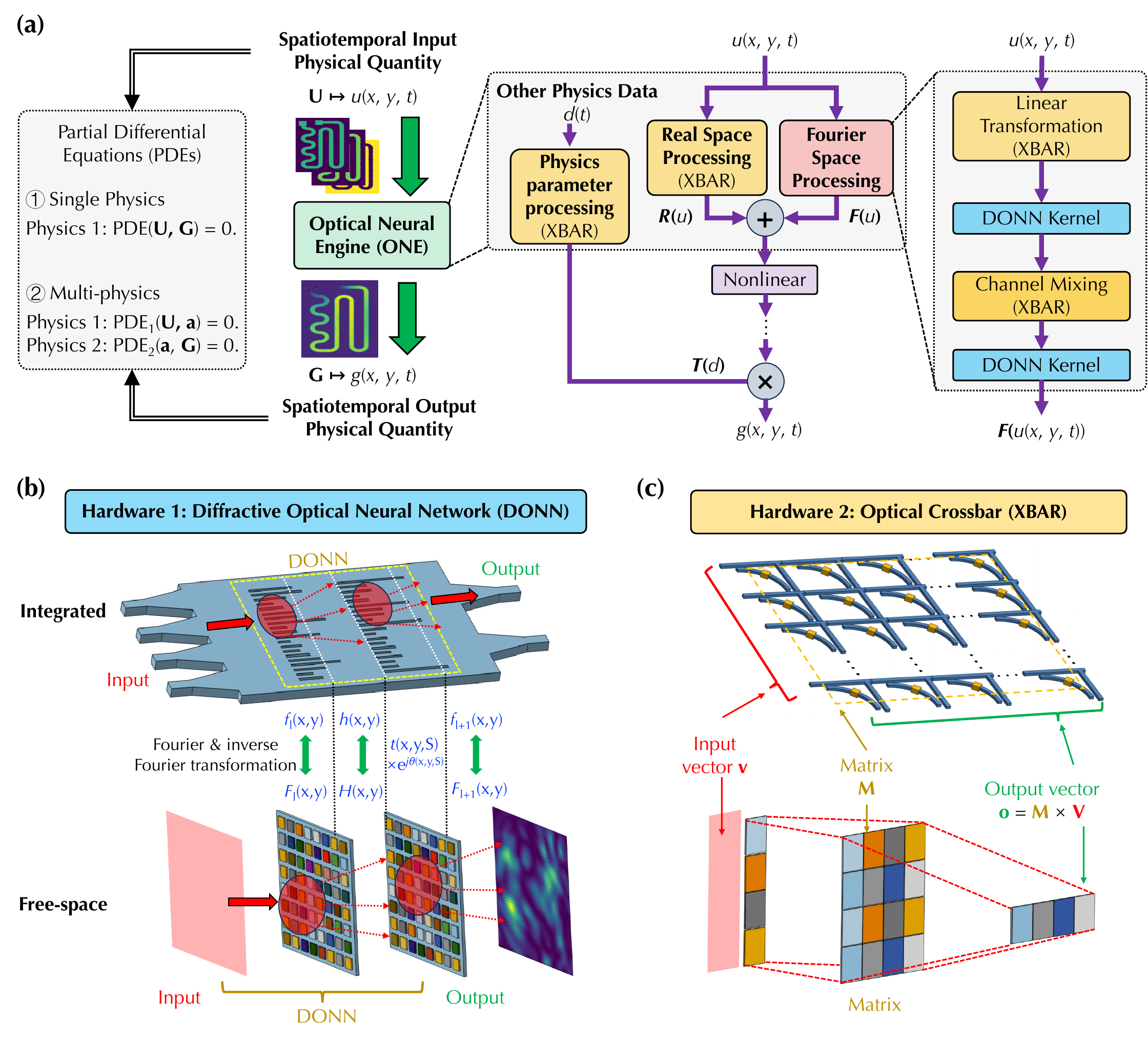}
    \vspace{-15pt}
    \caption{\textbf{ONE architecture and hardware implementations}. (a)~Illustration of processing branches and flows in the ONE architecture to predict output spatiotemporal output physical quantities from corresponding input and solve PDEs involving single or multiple physics. Illustrations of integrated and free-space implementations of reconfigurable (b)~DONN and (c)~XBAR structures.}
    \label{fig:ONE_arch}
\end{figure}

Figure\,\ref{fig:ONE_arch}a illustrates the ONE architecture, which takes the spatiotemporal data of an input physical quantity $\textbf{U}$, described as a function $u(x,y,t)$ in terms of positions $x$ and $y$ and time $t$, to predict the spatiotemporal data of an output physical quantity $\textbf{G}$ described using a function $g(x,y,t)$. The input and output quantities $\textbf{U}$ and $\textbf{G}$ can be connected through either a single-physics PDE or coupled multiphysics PDEs. There are three branches inside the ONE architecture, including (i)~Fourier space processing branch, (ii)~real space processing branch, and (iii)~physics parameter processing branch. The combination of both real and Fourier space processing has been proven fast, powerful, and efficient in solving PDEs~\cite{LiEtAl2020APA}, and the incorporation of additional physics parameter processing enables the fusion of multimodal data for complex tasks~\cite{LuEtAl2019APA}. More importantly, most operations in these branches can be deployed on optical computing hardware in both real and Fourier space, enabling solving PDEs in high-throughput and energy-efficient manners. The details of each branch are described below. 

In the first Fourier space processing branch, the core arithmetic operations are based on Fourier and inverse Fourier transformations to process input spatiotemporal data in the Fourier space. Their optical hardware implementations are mainly based on reconfigurable DONNs, which contain cascaded reconfigurable diffractive layers. Reconfigurable DONNs can be implemented in both integrated photonic chips~\cite{WangEtAl2022NC-2,ZhuEtAl2022NC} and free space~\cite{LinEtAl2018S,ZhouEtAl2021NP,ChenEtAl2022LPR}; see Fig.\,\ref{fig:ONE_arch}b. There are two fundamental operations in DONNs -- optical diffraction and spatial light modulation. For the optical diffraction operation, an optical field right after the $l$-th diffractive layer, $f_l$, diffracts to the front of $(l+1)$-th layer, whose optical field, $f_{\textrm{in},l+1}$, is a convolution of $f_l$ and the diffraction impulse function $h(x,y)$.  Specifically, the complex-valued field at point $(x,y)$ on the input plane of $(l+1)$-th layer can be written as the convolution of all fields at the output plane of $l$-th layer as
\begin{align*}
    f_{\textrm{in},l+1}(x,y,z)={\iint}f_l(x',y',0)h(x-x',y-y')dx'dy',
\end{align*}
where $z$ is the distance between two diffractive layers and $h(x,y)$ is the impulse response function of free space. By the convolution theorem, this 2D convolution can be efficiently calculated in Fourier space based on Fourier and inverse Fourier transformations. Specifically, the 2D Fourier transformation $\mathcal{F}_{xy}$ of $f$ and $h$, $F$ and $H$, are connected through
\begin{align*}
% \begin{split}
    \mathcal{F}_{xy}(f_{\textrm{in},l+1}(x,y,z))&=\mathcal{F}_{xy}(f_{l}(x,y,0))\mathcal{F}_{xy}(h(x,y)),\\
    F_{\textrm{in},l+1}(\alpha,\beta, z)&=F_{l}(\alpha,\beta, 0)H(\alpha,\beta),
% \end{split}
\end{align*}
where $\alpha, \beta$ are spatial domain indices. After diffraction, the 2D inverse Fourier transformation $\mathcal{F}^{-1}_{xy}$ of $F_{\textrm{in},l+1}(\alpha,\beta, z)$, $f_{\textrm{in},l+1}(x, y, z)$, is then spatially modulated. Each diffraction pixel at location $(x,y)$ has a complex-valued electric field transmission coefficient $t(x,y,S)e^{\phi(x,y,S)}$, where $t(x,y,S)$ ($\phi(x,y,S)$) is the amplitude (phase) response as a function of external stimuli $S$, such as voltages. The spatial light modulation operation is expressed as a pixel-wise multiplication
\begin{align*}
% \begin{split}
    f_{l+1}(x,y,z) &= \mathcal{F}^{-1}_{xy}(F_{\textrm{in},l+1}(\alpha,\beta, z))t(x,y,S)e^{\phi(x,y,S)}\\
        &= f_{\textrm{in},l+1}(x,y,z)t(x,y,S)e^{\phi(x,y,S)}, 
% \end{split}
\end{align*}
where $f_{l+1}(x,y,z)$ is the near-field output field right after the $(l+1)$-th layer. More details can be found in \emph{Methods}. 

Before and between DONN kernels, there is a linear transformation operation based on fully connected layers to scale up the number of channels and a channel mixing operation based on matrix multiplications~\cite{LiEtAl2020APA}. The core arithmetic operations are based on MVM. Their optical hardware implementations are mainly based on reconfigurable optical XBAR structures, which encode element values of vector $\mathbf{v}$ and matrix $\mathbf{M}$ into light intensity through electro-optic modulators, perform multiplications through cascaded modulators, and add signals at the output detector array. The signals are routed to follow mathematical calculations in MVM so that the reading from the detector array represents the output vector $\mathbf{o} = \mathbf{M}\times\mathbf{v}$. Reconfigurable XBAR structures can also be implemented in both integrated photonic chips~\cite{ShenEtAl2017NP,FeldmannEtAl2021N} and free space~\cite{HamerlyEtAl2019PR,WangEtAl2022NC,ChenEtAl2023NP}; see Fig.\,\ref{fig:ONE_arch}c. More details on the operation mechanism can be found in \emph{Methods} and Supplementary Fig.\,1. 

The second real space processing branch contains fully connected layers, whose operations are also based on MVM and implemented with optical XBAR structures. The output from the Fourier space branch,  \textbf{\textit{F}}($u$), and the output from the real space branch, \textbf{\textit{R}}($u$) are added and further processed with a nonlinear operation. Note that the nonlinear operation is the only operation performed in electronic hardware in the ONE architecture.  Moreover, this combination of real space, Fourier space, and nonlinear processing is scaled up, repeated four times, and cascaded in series. The third branch is to perform a linear transformation on other relevant physics parameters $d(t)$, which are time sequences instead of spatiotemporal data, based on fully connected layers. The obtained data \textbf{\textit{T}}($d$) is multiplied and merged onto two other branches to have the final output $g(x,y,t)$.  Hence, except nonlinear operations, all other operations can be done with DONN and optical XBAR systems. These two systems can be seamlessly assembled into a single integrated photonic chip or a single free-space optical system for all-optical operations without converting between optical and electronic hardware, fully leveraging the advantages of high throughput and high parallelism in optical computing systems. More details on the ONE architecture model are in \emph{Methods}.

\subsection*{Darcy flow and magnetostatic Poisson's equations}
\begin{figure}[hbt]
    \centering
    \includegraphics[width=\textwidth]{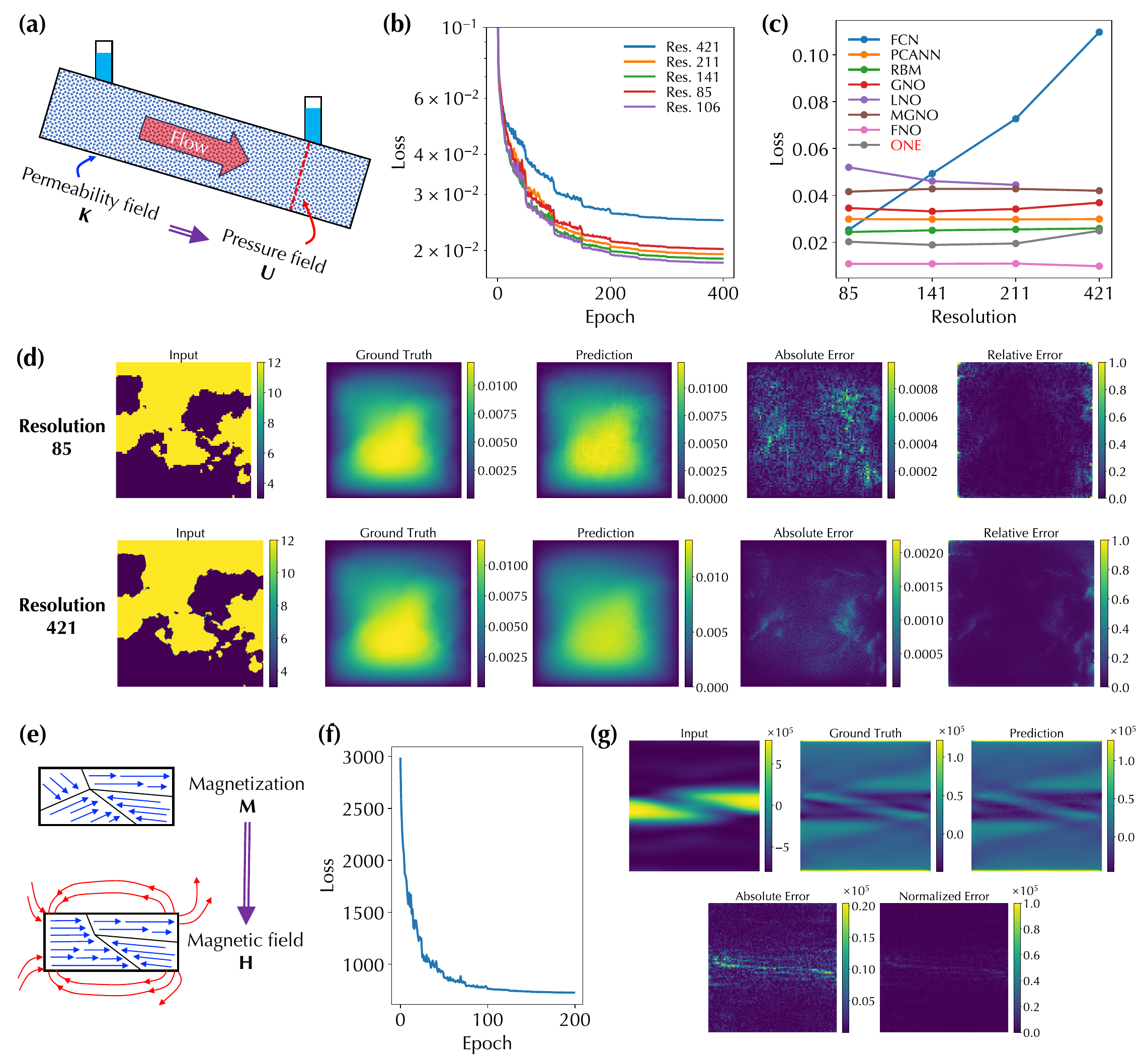}
    \vspace{-15pt}
    \caption{\textbf{Solving Darcy flow and magnetostatic Poisson's equations}. (a)~Illustration of the Darcy flow equation describing a fluid flow through a porous medium. The ONE architecture learns the mapping between the permeability and pressure fields. (b)~Training loss curves for input data with different resolutions. (c)~Comparison of the training loss of different models at various resolutions. (d)~Input permeability field, the expected ground truth of output pressure field, the predicted output pressure field, the absolute error between the expected and predicted outputs, and the relative error between the expected and predicted outputs, at $85$ and $421$ resolutions. (e)~Illustration of the magnetostatic Poisson's equation calculating the demagnetizing field generated by the magnetization field. The ONE architecture learns the mapping between these two fields. (f)~Validation loss curve for the ONE architecture solving the magnetostatic Poisson's equation and (g)~corresponding input magnetization field, the expected ground truth of output demagnetizing field, the predicted output demagnetizing field, the absolute and normalized errors between the expected and predicted outputs.}
    \label{fig:darcy}
\end{figure}

The first PDE we solved with the ONE architecture is the Darcy flow equation in fluid dynamics physics. This PDE describes a fluid flow through a porous medium as shown in Fig.\,\ref{fig:darcy}a. Specifically, the equation is
\begin{align*}
    -\nabla \cdot (k(x,y)\nabla u(x,y)) &= f(x,y),
\end{align*}
where $k(x,y)$ is the permeability field of the medium, $u(x,y)$ is the pressure field of the flow, and $f(x,y)$ is the force function. The ONE architecture was trained to learn the mapping from the 2D function $k(x,y)$ to function $u(x,y)$. More details about the equation dataset generation and training are in \emph{Methods}. Figure\,\ref{fig:darcy}b displays the training loss curves for inputs with different resolutions. The training loss is generally low for all resolutions and slightly increases at the highest $421$ resolution. Figure\,\ref{fig:darcy}c shows the comparison of the training loss of our ONE architecture with other PDE solving models, including fully convolution networks (FCN)~\cite{ZhuEtAl2018JCP}, principal component analysis-based neural network (PCANN)~\cite{BhattacharyaEtAl2021SJCM}, reduced biased method (RBM)~\cite{DeVoreEtAl2017}, graph neural operator (GNO)~\cite{LiEtAl2020APA-2}, low-rank kernel decomposition neural operator (LNO)~\cite{LuEtAl2019APA}, multipole graph neural operator (MGNO)~\cite{LiEtAl2020ANIPS}, and Fourier neural operator (FNO)~\cite{LiEtAl2020APA}. The performance of the ONE architecture is comparable with the state-of-the-art neural operators including GNO, LNO, MGNO, and FNO, and is better than FCN. Further, from the hardware perspective, the ONE architecture is constructed based on high-throughput optical computing hardware platforms so that all operations can be performed in parallel within a single clock cycle. In addition, the ONE architecture can be practically implemented on a large scale. For example, free-space reconfigurable DONNs~\cite{ZhouEtAl2021NP,ChenEtAl2022LPR,ChenEtAl2023N} and optical MVM~\cite{WangEtAl2022NC} are typically implemented using spatial light modulators (SLMs) with a scale $>1000\times 1000$. Hence, the execution cost of solving PDEs with different scales and resolutions is invariant, meaning $\mathcal{O}(1)$, if the scale of the optical hardware in the ONE architecture is large enough. Figure\,\ref{fig:darcy}d displays the input permeability field $k(x,y)$, the expected ground truth of output pressure field $u(x,y)$, the predicted output pressure field, the absolute error between the expected and predicted outputs, and the relative error between the expected and predicted outputs, at the lowest $85$ and the highest $421$ resolutions, respectively. This visualization further validates the ONE architecture in solving PDEs. More data on other resolutions are shown in Supplementary Fig.\,2. 

%%% Figure 2 %%%
%%%%%%%%%%%%%%%

The second PDE we solved is the magnetostatic Poisson's equation of demagnetization in micromagnetics physics. This PDE calculates the demagnetizing field $\mathbf{H}$ generated by the magnetization field $\mathbf{M}$ as shown in Fig.\,\ref{fig:darcy}e. Specifically, the equation is obtained from Maxwell's equation as
\begin{align*}
    \mathbf{\nabla} \cdot \mathbf{H} = -\mathbf{\nabla}\cdot \mathbf{M}.
\end{align*}
By defining an effective magnetic charge density $\rho = -\mathbf{\nabla}\cdot \mathbf{M}$ and a magnetic scalar potential $\Phi$ assuming there is no free current, we can express the demagnetizing field $\mathbf{H} = -\mathbf{\nabla}\Phi$ and rewrite the previous equation as a Poisson's equation
\begin{align*}
    \nabla^2\Phi = -\rho.
\end{align*}

Similar to solving the Darcy flow equation, the ONE architecture was trained to learn the mapping from components of $\mathbf{M}$ to $\mathbf{H}$ vector fields. More details about the equation dataset generation and training are in \emph{Methods}. Figure\,\ref{fig:darcy}f shows the validation loss curve and Fig.\,\ref{fig:darcy}g shows the input one component of $\mathbf{M}$ field, the expected ground truth of output $H_x$ component of $\mathbf{H}$ field, the predicted output $H_x$ component, the absolute error between the expected and predicted outputs, and normalized error between the expected and predicted outputs with respect to the maximum field strength in the ground truth. Both confirm a good performance of the ONE architecture in solving the magnetostatic Poisson's equation. More data on $H_y$ and $H_z$ components is shown in Supplementary Fig.\,3. 

\subsection*{Navier-Stokes and Maxwell's equations}

\begin{figure}[hbt]
    \centering
    \includegraphics[width=\textwidth]{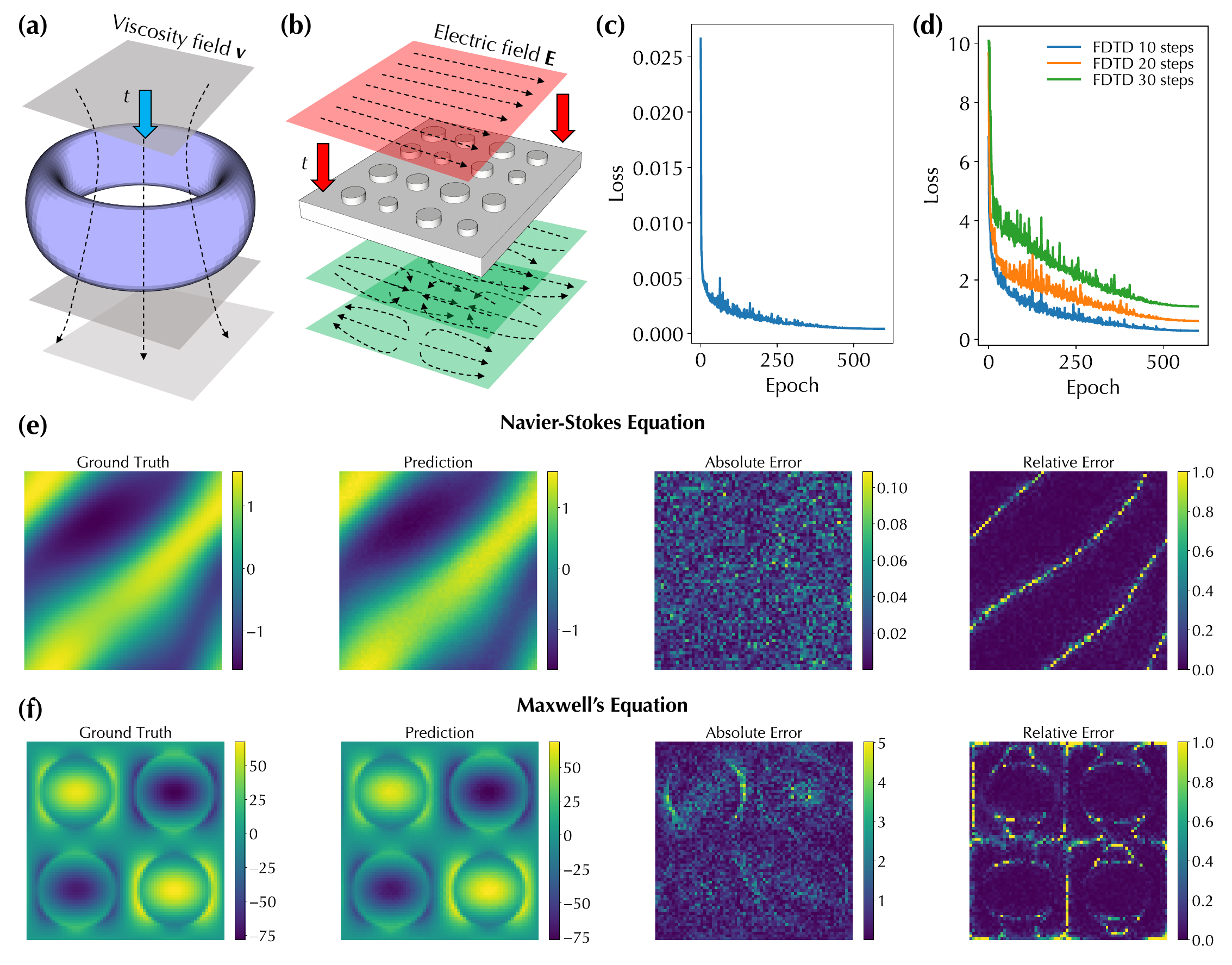}
    \caption{\textbf{Solving time-dependent Navier-Stokes and Maxwell's equations}. Illustrations of (a)~Navier-Stokes equation for solving the time evolution of the vorticity field in a viscous, incompressible fluid in vorticity form on the unit torus and (b)~Maxwell's equations for solving the time evolution of the electric field in a dielectric metasurface. Validation loss curves for (c)~solving the Navier-Stokes equation and (d)~Maxwell's equations using the ONE architecture. The expected ground truth field, the predicted field, and the absolute and relative errors between these two fields for (e)~the Navier-Stokes equation and (f)~Maxwell's equations, respectively.}
    \label{fig:ns_fdtd}
\end{figure}

In addition to steady-state Darcy flow and magnetostatic Poisson's equations without time evolution, we employed the ONE architecture to solve time-dependent PDEs, including the Navier-Stokes equation in fluid dynamics and Maxwell equations in electromagnetics and optics. In particular, the real-time reconfigurability of DONN and optical XBAR structures makes the ONE architecture suitable for such a purpose. Specifically, we solved a 2D Navier-Stokes equation for a viscous, incompressible fluid in vorticity form on the unit torus as shown in Fig.\,\ref{fig:ns_fdtd}a. This PDE calculates the time evolution of vorticity described as
\begin{align*}
\partial_t w(x,y,t) + u(x,y,t)\cdot\nabla w(x,y,t) &= v\Delta w(x,y,t) + f(x,y), 
\end{align*}
where $u$ is the velocity field, $w = \nabla\times u$ is the vorticity, $\nu$ is the viscosity coefficient, $f$ is the forcing function. The ONE architecture was trained to learn the mapping from $w$ in a time range from 0 to $t_0$ to $w$ in a time range from $t_0$ to $t_1$ ($t_1>t_0$). More details about the equation dataset generation and training are in \emph{Methods}. Further, we also solved Maxwell's equations in a dielectric metasurface consisting of multiple cylindrical pillars in a unit cell of a periodic pattern as shown in Fig.\,\ref{fig:ns_fdtd}b~\cite{TangEtAl2022NCS}. The general Maxwell's equations can calculate the time evolution of an electric field through the following equations
\begin{align*}
  \nabla \cdot \mathbf{D} &= \rho,\\
  \nabla \cdot \mathbf{B} &= 0,\\
  \nabla \times \mathbf{E} &= -\frac{\partial \mathbf{B}}{\partial t},\\
  \nabla \times \mathbf{H} &= \mathbf{J} + \frac{\partial \mathbf{D}}{\partial t}, 
\end{align*}
where $\mathbf{D}$ is the electric displacement field, $\rho$ is the free charge density, $\mathbf{B}$ is the magnetic flux density, $\mathbf{E}$ is the electric field, $\mathbf{H}$ is the magnetic field, and $\mathbf{J}$ is the free current density. The ONE architecture was trained to learn the mapping from $\mathbf{E}$ in a time range from 0 to $t_0$ to $\mathbf{E}$ in a time range from $t_0$ to $t_1$ ($t_1>t_0$). More details about the dataset generation and training are in \emph{Methods}. Figure\,\ref{fig:ns_fdtd}c displays the validation loss curve for solving the Navier-Stokes equation with $t_0 = 10$ and $t_1 = 20$. Figure\,\ref{fig:ns_fdtd}d displays the validation loss curves for solving Maxwell's equations with $t_0 = 10$ and $t_1 = 20, 30, 40$, respectively. Moreover, Figure\,\ref{fig:ns_fdtd}e and \ref{fig:ns_fdtd}f show the expected ground truth of $w$ field and the $E_x$ component of the $\mathbf{E}$ field at $t_1$, the corresponding predicted fields at $t_1$, and the absolute and relative errors between ground truth and prediction for the Navier-Stokes equation and Maxwell's equations, respectively. All confirm a good performance in solving time-dependent PDEs using the ONE architecture. 

\subsection*{Multiphysics PDEs}

\begin{figure}[hbt]
    \centering
    \includegraphics[width=\textwidth]{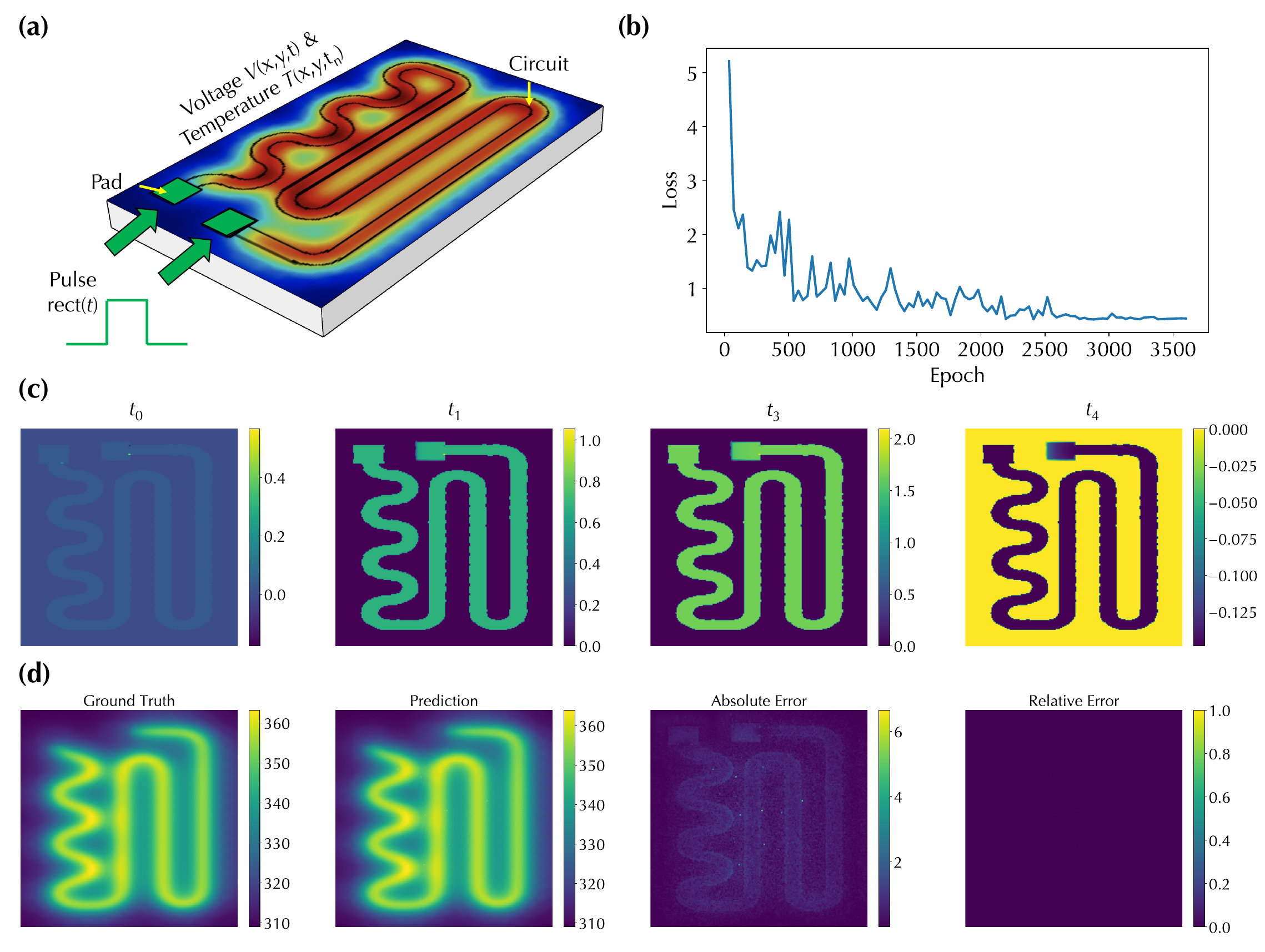}
    \caption{\textbf{Solving multiphysics PDEs}. (a)~Illustration of solving coupled PDEs in an electrical heating problem involving electric current physics and heat transfer physics. (b)~Validation loss curve. (c)~A few representative 2D voltage profiles in the circuit. (d)~The expected ground truth temperature profile, the predicted profile, and the absolute and relative errors between these two profiles.}
    \label{fig:comsol}
\end{figure}

Moreover, we employed the ONE architecture to solve coupled PDEs involving two physics. Specifically, we solved an electrical heating problem to obtain a temperature profile at an intermediate time step $t_n$, $T(x,y, t_n)$, in an electrical circuit when a time-dependent voltage signal was applied to the circuit pads, involving coupled electric current physics and heat transfer physics; see Fig.\,\ref{fig:comsol}a. Specifically, for the electrical current physics, the corresponding PDE is
\begin{align*}
Q_e &= d\sigma\nabla_t V(x,y,t), \\
V(x_0, y_0, t) &= \textrm{rect(t)},
\end{align*}
where $Q_e$ is the heat rate per unit area from an electromagnetic heating source, $d$ is the thickness of the heating layer, $V(x,y,t)$ is the voltage profile in the circuit that is subjected to a voltage boundary condition defined in the pads $V(x_0, y_0, t)$, and $V(x_0, y_0, t)$ is a pulse rectangular function rect($t$) with pulse height and width. For the heat transfer physics, the corresponding PDE is
\begin{align*}
\rho C_p \frac{\partial T}{\partial t} + \rho C_p \mathbf{u} \cdot \nabla T - \nabla \cdot (k\nabla T) &= Q_e,
\end{align*}
where $\rho$ is the mass density, $C_p$ is the specific heat capacity, $T$ is the absolute temperature, and $k$ is the thermal conductivity. These two PDEs are connected through the quantity $Q_e$. The ONE architecture was trained to learn the mapping from $V(x,y,t)$ in a time range spanning all time steps in input pulses to $T(x,y, t_n)$ at an intermediate pulse time step $t_n$. In contrast to previous examples, the pulse information, including pulse height and width, was processed through the physics parameter processing branch in the ONE architecture (Fig.\,\ref{fig:ONE_arch}a) and multiplied with the output from cascaded real space processing and Fourier space processing branches to yield the final output. More details about the dataset generation and training are in \emph{Methods}. Figure\,\ref{fig:comsol}b displays the validation loss curve and Fig.\,\ref{fig:comsol}c shows a few representative input 2D data $V(x,y,t)$ at various time steps. Figure\,\ref{fig:comsol}d shows the expected ground truth of $T(x,y,t_n)$, the corresponding predicted temperature profile, and the absolute and relative errors between ground truth and prediction. All confirm a good performance in solving multiphysics PDEs using the ONE architecture. 

\subsection*{Experimental demonstration}

\begin{figure}[hbt]
    \centering
    \includegraphics[width=\textwidth]{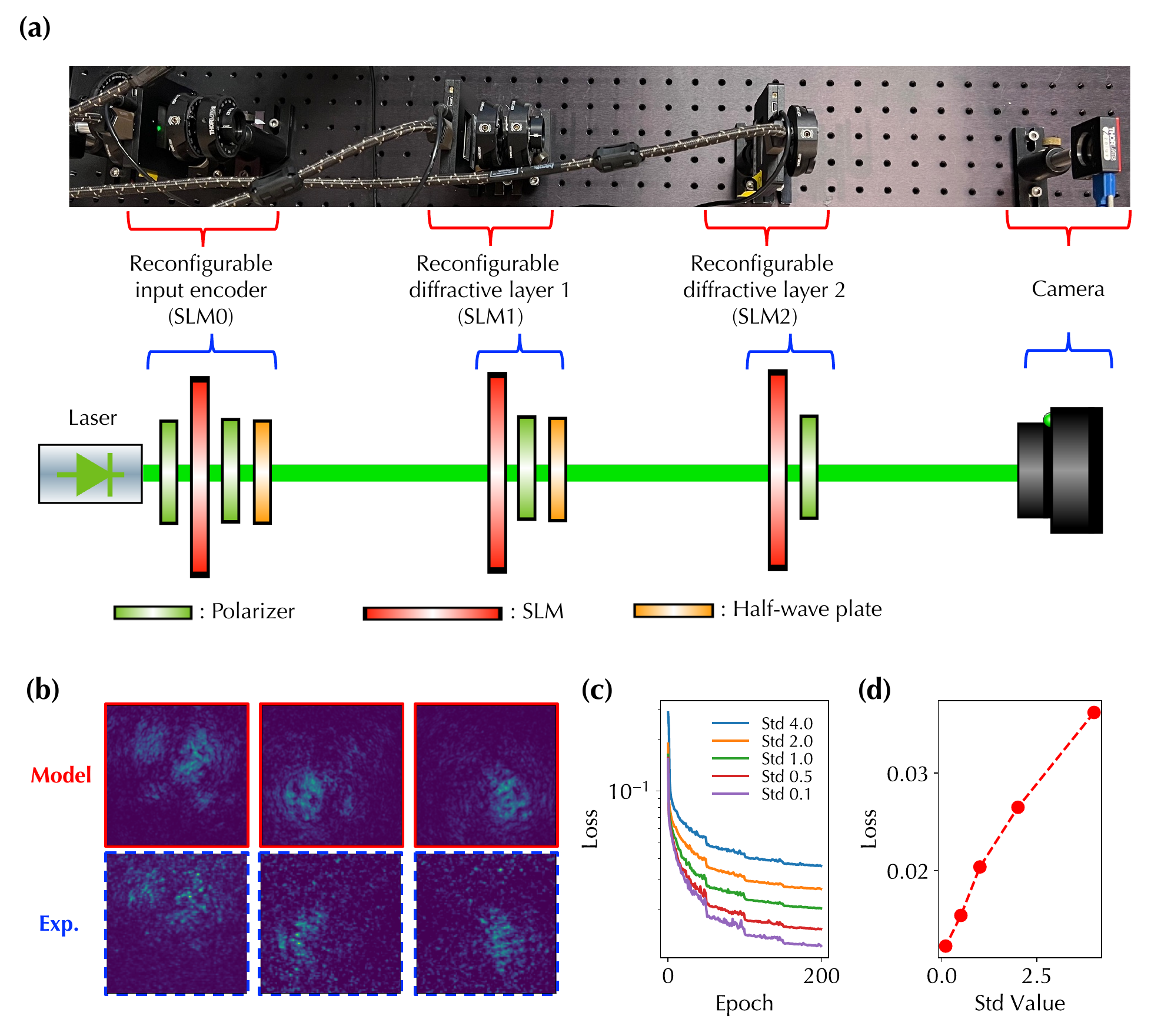}
    \caption{\textbf{Experimental demonstration}. (a)~Photo and schematic of a reconfigurable DONN experimental setup consisting of a reconfigurable input encoder, two reconfigurable diffractive layers, and a camera. Polarization components were used to configure SLMs in the phase modulation mode. (b)~Output 2D data in one DONN kernel of the Fourier space processing branch in the ONE architecture obtained from model calculations and experimental measurements. (c)~Validation loss curves at different noise levels in optical XBAR structures and (d)~the loss at the final epoch as a function of noise level. }
    \label{fig:exp}
\end{figure}

Finally, to demonstrate the experimental feasibility of the ONE architecture, we constructed a free-space reconfigurable DONN setup and evaluated the performance of solving the Darcy flow equation under different hardware noise levels in optical XBAR structures. Figure\,\ref{fig:exp}a displays a photo and schematic of the reconfigurable DONN setup, which contains a laser source, a reconfigurable input encoder, two reconfigurable diffractive layers, and a camera. The reconfigurable encoder and diffractive layers were built upon SLMs, which can modulate the amplitude and phase of transmitted light when applying voltage. Multiple light polarization components, including polarizers and half-wave plates, were also employed to manipulate polarization states to achieve large phase modulation ranges. More details on the experimental setup are in \emph{Methods}. 

As shown in Supplementary Fig.\,4, the experimentally measured amplitude and phase modulation responses of all three SLMs are not only discrete with respect to grey levels but also coupled and dependent. To leverage the gradient-based ML training algorithm, we utilized the \texttt{Gumbel-softmax} reparameterization technique to approximate a discrete distribution to a continuous distribution~\cite{ChenEtAl2022LPR}. More details are described in \emph{Methods}. Moreover, the values of input 2D data span both negative and positive values and were encoded as the grey level of the SLM in the reconfigurable input encoder (SLM0 in Fig.\,\ref{fig:exp}a). We performed the encoding through linear mapping from minimum and maximum values of input data to a grey-level range in the SLM. More details are described in \emph{Methods}. In addition, we precisely aligned all SLMs with respect to each other within a range of a few pixels on the order of hundreds of $\mu$m; see Supplementary Fig.\,5. Although the long optical path in the system makes the alignment sensitive to external variations, the system's full reconfigurability can enable fast adaptive pixel-by-pixel re-alignment. Figure\,\ref{fig:exp}b shows output 2D data in one DONN kernel of the Fourier space processing branch in the ONE architecture (Fig.\,\ref{fig:ONE_arch}a) obtained from model calculations and experimental measurements, showing good agreement and experimentally validating the feasibility of the ONE architecture in solving PDEs. More data is shown in Supplementary Fig.\,6. There are some speckles in the background of measured images, which probably originate from high-order diffraction interference, leading to numerical errors in the ONE architecture for performing regression tasks. This discrepancy between models and experiments can be mitigated through hardware-software co-design, such as incorporating loss functions based on experimental results for gradient calculations as demonstrated in prior works~\cite{ZhouEtAl2021NP,WrightEtAl2022N,ChenEtAl2023N}. 

We also evaluated the performance of the ONE architecture under different noise levels of optical XBAR structures. Specifically, we added random Gaussian noise with zero mean and varying standard deviation (Std) to the values obtained from matrix multiplications to represent hardware noise, such as shot noise in photodetectors~\cite{FanEtAl2023AIS}. The corresponding MVM results and histograms of different noise standard deviation values are shown in Supplementary Fig.\,7, and more details can be found in \emph{Methods}. As shown in Fig.\,\ref{fig:exp}c and Fig.\,\ref{fig:exp}d, the validation loss increases with the increasing noise standard deviation value. The current hardware implementation of optical XBAR structures with advanced components and calibration algorithms~\cite{HamerlyEtAl2019PR,WangEtAl2022NC,ChenEtAl2023NP}, including the structure we demonstrated before~\cite{FanEtAl2023AIS}, can achieve quite a small noise level similar or below the noise level corresponding to 0.5 Std. Hence, the noise influence in optical XBAR structures on the performance of the ONE architecture is not substantial. 

We further estimated the potential throughput and power consumption of the ONE architecture implemented using optical computing hardware for inference. The throughput is mainly determined by the SLM refresh rate and camera frame rate. Current commercial SLMs and cameras can have rates $>1000\,$Hz, meaning that the inference time for one instance is $<1\,$ms. In contrast, it typically takes minutes to hours to numerically solve PDEs. Hence, the ONE architecture features $>10^5$ (five orders of magnitude) acceleration compared to typical PDE solvers. This throughput is also comparable to the state-of-art ML model, such as FNO with a 5\,ms inference time~\cite{LiEtAl2020APA}. Moreover, the system throughput can be substantially improved with device innovation. For example, an electro-optic SLM based on organic molecules can achieve $>$\,GHz switching speed~\cite{Benea-ChelmusEtAl2022NC}, and an ultrafast camera can achieve a trillion frames per second~\cite{KimEtAl2020SA}. With these devices, the ONE architecture can achieve an inference time $<1\,$ns. The power consumption is mainly determined by the leakage current of liquid crystal cells in SLMs. Because of the dielectric nature of liquid crystals and their high leakage resistance, the leakage current is typically $<1\,\mu$A. Hence, assuming a 10\,V driving voltage, the static power consumption of SLMs is $\sim10\,\mu$W, which is nearly $10^7$ (seven orders of magnitude) smaller than typical GPU inference power $\sim100\,$W.  

\section*{Discussion}

We have demonstrated the ONE architecture and validated its performance in solving a broad range of PDEs in diverse scientific domains. The ONE architecture is versatile and can be modified to reduce the interface and connection between DONN and optical XBAR structures and facilitate the hardware implementation of the whole system. Further, in a whole system, active learning and noise-aware training can be incorporated to mitigate the discrepancy between models and practical systems for accurate deployment. Moreover, in addition to solving PDEs, the ONE architecture can be tailored to accelerate ML models for other regression problems. 

\section*{Methods}

\smallskip
\noindent\textbf{DONN diffraction model} -- The diffraction impulse function $h(x,y)$ was described using the Fresnel equation as
\begin{align*}
    h(x,y)=\frac{e^{ikz}}{i\lambda z}e^{\frac{ik}{2z} (x^2+y^2)}, 
\end{align*}
where $\lambda$ is the wavelength, $k = 2\pi/\lambda$ is the free-space wavenumber, $(x,y)$ are positions within a plane perpendicular to the wave propagation direction, $z$ is the distance along the propagation direction, and $i$ is the imaginary unit. The 2D Fourier transformation was directly performed on $h(x,y)$ for model training and evaluation. To match the experimental setup as described below, $h(x,y)$ was first discretized with respect to a defined rectangular mesh grid in the convolution calculation and then converted into the Fourier space through 2D Fourier transformation. More details can be found in our prior work~\cite{ChenEtAl2022LPR}.

\smallskip
\noindent\textbf{The operation mechanism of optical XBAR structures} -- Supplementary Fig.\,1a shows the detailed schematic of an integrated photonic XBAR structure. Specifically, the element values of a $n \times 1$ input vector $\mathbf{v}$ are represented by the intensities of light at input waveguides, $\{I_1, I_2, I_3, ..., I_n\}$, which can be implemented by modulating an equally distributed laser intensity through a $n \times 1$ array of electro-optic modulators (red squares in Supplementary Fig.\,1a) at input waveguides. The light on each row waveguide is then equally distributed to the column waveguides connected to that row waveguide and modulated through an electro-optic modulator on the coupled curved waveguide (yellow squares in Supplementary Fig.\,1a). The element values of a $ m\times n$ matrix $\mathbf{M}$ are represented by the transmittance of modulators on curved waveguides, $\{T_{ij}\}, i\in[1,m], j\in[1,n]$. At the end of each column waveguide, a photodetector collects all light intensity passing through the column waveguide. The obtained photocurrents or photovoltages of a $m \times 1$ photodetector array represent the summation of multiplied input vector light intensity and matrix modulator transmittance, and the element values of output vector $\mathbf{o}$, $O_{j} = \sum_{s=1}^{n}T_{js}I_{s}, j\in [1,m]$. Hence, this integrated photonic XBAR structure can implement MVM in the optical domain. 

Similarly, Supplementary Fig.\,1b shows the detailed schematic of a free-space optical XBAR structure. Specifically, the element values of a $n \times 1$ input vector $\mathbf{v}$ are represented by the intensities of light, $\{I_1, I_2, I_3, ..., I_n\}$, which is implemented through a $n \times 1$ array of free-space vector SLM. The output light is broadcast to a $ m\times n$ array of matrix SLM through lenses so that the light distribution from vector SLM is identical at each column of matrix SLM. The element values of a $ m\times n$ matrix $\mathbf{M}$ are represented by the transmittance of matrix SLM, $\{T_{ij}\}, i\in[1,m], j\in[1,n]$. Lenses are then used to focus the output light from each modulator on the same column of matrix SLM to a photodetector. The readings from a $m \times 1$ photodetector array represent the element values of output vector $\mathbf{o}$, $O_{j} = \sum_{s=1}^{n}T_{js}I_{s}, j\in [1,m]$. Hence, this free-space optical XBAR structure can also implement MVM in the optical domain. 

\smallskip
\noindent\textbf{ONE architecture model} -- The ONE architecture model was constructed with two main modules -- the DONN module processing data in the Fourier space and the optical XBAR module processing linear operations. The mathematical operations in DONN and optical XBAR structures have been described before and their accurate models have been implemented in our prior works, closely matching experimental results ~\cite{ChenEtAl2022LPR,FanEtAl2023AIS}. Briefly, the DONN module was modeled by combining the Fresnel free-space diffraction with phase-only spatial light modulation in a range of $[0,2\pi]$ in the model and coupled spatial light modulation as shown in Supplementary Fig.\,4; the optical XBAR module was represented as matrix multiplication incorporating measurement noise. Both modules were implemented under the \texttt{PyTorch 1.12} framework with graphics processing unit (GPU)-accelerated parallel computation and gradient backpropagation for training. The GPU used in this work was an Nvidia RTX 6000 card. 

\smallskip
\noindent\textbf{Darcy flow equation dataset and training} -- A 2D Darcy flow equation on the unit box was employed as described in detail in Ref.\,~\cite{LiEtAl2020APA}. The corresponding PDE is a second-order, linear, elliptic PDE as
\begin{align*}
-\nabla \cdot (k(x,y)\nabla u(x,y)) &= f(x,y),  &x\in(0,1), y\in (0,1),\\
u(x) &= 0, &x\in\partial(0,1), y\in\partial(0,1)
\end{align*}
with a Dirichlet boundary condition. We used the Darcy flow dataset from the existing dataset in Ref.\,\cite{LiEtAl2020APA} with a boundary condition $u(x,y)=0$ on domain edges. The coefficient $k(x,y)$ was generated based on a specific distribution with the value 12 for positive inputs and 3 for negative inputs. The forcing term was fixed at $f(x,y)=1$. The solution $u(x,y)$ was computed using a second-order finite difference method on a 421 × 421 grid, and other resolutions were obtained with downsampling. We used a $10:1$ ratio for the numbers of data in the training set and validation set, respectively. The model was trained with a total of 600 epochs and a batch size of 40. The learning rate was 0.1 for the trainable parameters in DONNs and 0.001 for all other trainable parameters with the Adam optimizer.

\smallskip
\noindent\textbf{Magnetostatic Poisson's equation dataset and training} -- The demagnetizing field $\mathbf{H}$ originates from the magnetization within the material itself, which can be calculated as the convolution of $\mathbf{M}$ with the demagnetization tensor $\mathbf{N}$ as
\begin{align*}
    \mathbf{H(r)} = \int \mathbf{N(r-r')}\mathbf{M(r')}d\mathbf{r'}.
\end{align*}
This convolution was computed through Fourier space representations of fields. Specifically, to create the dataset, we utilized the MagneX solver~\cite{github_magnex} to simulate the time evolution of magnetization in a thin magnetic film with dimensions of $500\times 125\times 3.125$~nm. The modeling incorporated both demagnetization and exchange interactions. Initially, we relaxed the magnetic field into a stable S-state before subjecting the system to varying external magnetic fields in different scenarios. We uniformly sampled 8 bias $\mathbf{H}$ fields in the $x$ and $y$ directions, each with a magnitude of 19872 A/m. The system evolved for 1\,ns, during which we collected paired data of $\mathbf{M}$ and $\mathbf{H}$ fields. Each field was represented by three channels corresponding to the field components in $x$, $y$, and $z$ directions. The dataset was divided into training and testing sets with an $8:2$ ratio. The training was conducted over 500 epochs with a batch size of 128. The learning rate was set to 1.0 for the trainable parameters in DONNs and 0.001 for all other trainable parameters with the Adam optimizer.

\smallskip
\noindent\textbf{Navier-Stokes equation dataset and training} -- A 2D Navier-Stokes equation for a viscous, incompressible fluid in vorticity form on the unit torus was used to generate spatiotemporal data for training the ONE architecture. The details are described in Ref.\,\cite{LiEtAl2020APA}. Specifically, the PDEs are
\begin{align*}
\partial_t w(x,y,t) + u(x,y,t)\cdot\nabla w(x,y,t) &= v\Delta w(x,y,t) + f(x,y), & x\in(0,1), y\in(0,1), t\in(0,T]\\
\nabla\cdot u(x,y,t) &=0, & x\in(0,1), y\in(0,1), t\in(0,T]\\
w(x,y,0) &= w_0(x,y), & x\in(0,1), y\in(0,1),
\end{align*}
where $w_0(x,y)$ is the initial vorticity and boundary conditions were used. We utilized the existing dataset with the viscosity coefficient $v = 10^{-3}$ from Ref.\,\cite{LiEtAl2020APA} for training and inference. The samples in the dataset were recorded with a time step of $10^{-4}$\,s. We used 1000 data as the training set and 100 data as the validation set. We trained the ONE architecture model with the first 10 vorticity fields ($w(x,y,t)$) to predict the time evolution of the next 10 vorticity fields. The model was trained with a total of 600 epochs and a batch size of 40. The learning rate was 0.1 for the trainable parameters in DONNs and 0.001 for all other trainable parameters with the Adam optimizer. 

\smallskip
\noindent\textbf{Maxwell's equations dataset and training} -- We employed commercial Ansys Lumerical finite-difference-time-domain simulation software to generate an electric field dataset by solving Maxwell's equations in dielectric metasurfaces. Specifically, the dielectric metasurface had a periodic pattern and we used four silicon cylindrical rods as the unit cell and periodic boundary condition. Data were generated by randomly selecting the radii of four cylindrical rods. The radius was chosen from $39.5\,\mu$m to $44.5\,\mu$m with a step of $0.25\,\mu$m. The simulation time was set as 300000\,fs. We generated a total of 1200 data and used 1000 as the training set and the rest 200 as the validation set. The model was trained in an auto-regressive style for the $E_x$ component processing. The $E_x$ field data between 300000\,fs to 160000\,fs was backward fed into to the model to predict the next 40000\,fs $E_x$ field data. The model was trained with a total of 500 epochs and a batch size of 20. The learning rate was 0.1 for the trainable parameters in DONNs and 0.001 for all other trainable parameters with the Adam optimizer.

\smallskip
\noindent\textbf{Multiphysics dataset and training} -- We employed commercial COMSOL Multiphysics finite-element simulation software to generate a temperature profile dataset by solving coupled electric current and heat transfer PDEs in an electrical heating circuit. The circuit details can be found in Ref.\,\cite{comsol_heating}. Concisely, the circuit contained a serpentine-shaped Nichrome resistive layer with 10\,$\mu$m thick and 5\,mm wide on top of a glass plate. A silver contact pad with a dimension 10\,mm $\times$ 10\,mm $\times$ 10\,$\mu$m was attached at each end. The deposited side of the glass plate was in contact with the surrounding air at 293.15\,K and the back side was in contact with the heated fluid at 353\,K. Two coupled physics modules, electrical current in layered shells and heat transfer in layered shells, were used in COMSOL simulations. The input voltage pulse height was set from 5 to 25\,V with a step of 1\,V and the pulse width was set from 20 to 60\,s with a step of 1\,s. The simulation time range was from 0 to 110\,s. We generated a total number of 861 data and divided the data into training and testing set with the splitting ratio of $8:2$. The ONE architecture took the electric current layer data as the input spatiotemporal data and the input voltage pulse information was fed into the physics parameter data processing branch to predict temperature field data at 55\,s. The model was trained with a total of 100 epochs and a batch size of 40. The learning rate for the trainable parameters in DONNs was 0.1 and the learning rate for all other trainable parameters was 0.001 with the Adam optimizer.

\smallskip
\noindent\textbf{DONN experimental setup and alignment} -- The photo and schematic diagram of the DONN experimental setup are displayed in Fig.\,\ref{fig:exp}a. The laser diode with a center wavelength $532\,$nm (CPS532 from Thorlabs, Inc.) was used as a source. The distance between SLMs and between the last SLM and camera was set as $25.4\,$cm. The polarizers and half-wave plates before and after each SLM were configured so that each SLM operated with a strong modulation of the transmitted electric field phase (phase mode) together with a moderate modulation of light amplitude. The experimentally measured amplitude and phase modulation responses of three SLMs are shown in Supplementary Fig.\,4. All transmissive SLMs are the LC 2012 model from HOLOEYE Photonics AG with a refresh rate of 60\,Hz. The analog-to-digital converter has $8$-bit precision for liquid crystal driving voltage, so that the grey level of SLMs is from $0$ to $255$. The pixel size of SLMs is $36\,\mu$m$\times36\,\mu$m. The output data was captured on a CMOS camera with a frame rate of 34.8 frames per second (CS165MU1 from Thorlabs, Inc.). 

We aligned the DONN setup by loading standard images on SLMs and comparing experimental results with simulation. Specifically, as shown in Supplementary Fig.\,5a, standard Gaussian images, which were centered with a peak at 255 grey level and with a standard deviation of 6 pixels, were loaded in the input SLM and two diffractive SLMs. Supplementary Fig.\,5b displays the simulation pattern for the perfectly aligned setup. During the alignment process, loaded images were moved up, down, left, and right pixel-by-pixel to match the captured images by the camera with the simulation pattern. Supplementary Fig.\,5c displays the matched experimental diffraction pattern when the optical setup was aligned, while Supplementary Fig.\,5d shows misaligned patterns when there was five-pixel misalignment in vertical and horizontal directions, respectively. 

\smallskip
\noindent\textbf{DONN experimental training with reparameterization} --  The discrete look-up tables of device responses shown in Supplementary Fig.\,4 break the gradient backpropagation in the ML training process in \texttt{PyTorch}. To solve this challenge, we utilized a differentiable reparameterization \texttt{Gumbel-softmax} technique, which was first introduced in Ref.\,\cite{JangEtAl2016APA} and demonstrated in our prior work~\cite{ChenEtAl2022LPR}. Specifically, continuous noise from the Gumbel distribution was added to the discrete distribution. The \texttt{argmax} function was then used to find the optimized sample. The training problem after this \texttt{Gumbel-argmax} process is mathematically equivalent to the original training problem under one-hot representation~\cite{JangEtAl2016APA}. Since the \texttt{argmax} function still breaks the gradient chain, it was replaced with the \texttt{softmax} function to enable differentiability. Hence, this \texttt{Gumbel-softmax} technique, which is also available in \texttt{PyTorch}, offers continuous and differentiable approximation to discrete distributions and the gradient can backpropagate to reduce the loss function.

\smallskip
\noindent\textbf{DONN experimental grey-level encoding} -- The global minimum and maximum values in input 2D data were calculated as $d_{\textrm{min}}$ and $d_{\textrm{max}}$. A grey level range from $130$ to $255$ in the input encoder SLM was selected for a relatively large amplitude modulation range to have enough contrast. Hence, any value $d$ in the input 2D data was converted into a grey level through a linear mapping as
\begin{align*}
    d = \textrm{int}\left( \frac{ 255 - 130 }{d_{\textrm{max}} - d_{\textrm{min}}} + 130  \right),
\end{align*}
where the $\textrm{int}(\cdot)$ operation rounded the expression to the nearest integer since the SLM grey level must be an integer. 

\smallskip
\noindent\textbf{Optical XBAR noise} -- The MVM results from an optical XBAR structure were uniformly randomly generated in a range of $-15$ to $15$, which was the value range in the ONE architecture for solving the Darcy flow equation. The expected number $o$ was then added with a randomly generated noise from a Gaussian distribution with a zero average and varying standard deviation. The noise-dressed number $\tilde{o}$ was used in ONE architecture calculations. Under different noise standard deviation levels, Supplementary Fig.\,7a demonstrates $\tilde{o}$ with respect to $o$ and Supplementary Fig.\,7b displays histograms of $\tilde{o} - o$. 

\section*{Data availability}
Upon publication, all data that support the plots within this paper and other findings of this study will be available on a public \emph{GitHub} repository.  

\section*{Code availability}
Upon publication, all codes that support the plots within this paper and other findings of this study will be available on a public \emph{GitHub} repository. 

\section*{Acknowledgements}
R.C., C.Y., and W.G.\ acknowledge support from the National Science Foundation through Grants No.\ 2235276, No.\ 2316627, and No.\ 2428520. M.L., J.F., and W.G.\ also acknowledge support from the University of Utah start-up fund. Y.T., Z.Y., and A.N. were supported by Laboratory Directed Research and Development (LDRD) funding from Berkeley Lab, provided by the Director, Office of Science, of the U.S. Department of Energy under Contract No.\ DE-AC02-05CH11231.  This research used resources of the National Energy Research Scientific Computing Center (NERSC), a DOE Office of Science User Facility supported by the Office of Science of the U.S. Department of Energy under Contract No.\ DE-AC02-05CH11231 and under NERSC GenAI award under No.\ DDR-ERCAP0030541.

\section*{Author Contributions Statement}
Y.T.\ and W.G.\ conceived the idea and W.G.\ supervised the project. Y.T.\ constructed models and performed machine learning calculations with the help of M.L., J.F., and C.Y and under the support of A.N., Z.Y, and W.G. R.C\ constructed an optical experimental setup, performed experiments, and performed numerical calculations under the supervision of W.G. Y.T.\ and W.G.\ wrote the manuscript. 

\section*{Competing Interests Statement}
The authors declare no competing interests.


\end{document}